\documentclass[aps,prl,reprint,twocolumn,showpacs,floatfix,superscriptaddress]{revtex4-1}

\usepackage{amssymb,amsmath,amstext}                %%   American Physical Society math etc extensions
\usepackage{graphicx}                                               %%   Include figure files                                            
\usepackage{epstopdf}                                               %%   help with eps -> pdf 
\usepackage{color}                                                     %%   change color
\usepackage{bm}                                                        %%   bold math                                    
\usepackage{appendix}                                              %%   formating appendicies
\usepackage[utf8]{inputenc}
\usepackage{bbold}
\usepackage{bbm}
\usepackage{latexsym}
\usepackage{xcolor}
\usepackage{braket}
\definecolor{lblue} {RGB}{51,71,158}
\usepackage[colorlinks=true,citecolor=blue,linkcolor=blue,urlcolor=lblue]{hyperref}

\newcommand{\average}[1]{\overline{#1}}

\begin{document}

\title{Coexistence of localized and extended phases: \\ Many-body localization in a harmonic trap}
% Force line breaks with \\

\author{Titas Chanda}
\email{titas.chanda@uj.edu.pl}
\affiliation{Institute of Theoretical Physics, Jagiellonian University in Krak\'ow,  \L{}ojasiewicza 11, 30-348 Krak\'ow, Poland }

\author{Ruixiao Yao}
\affiliation{School of Physics, Peking University, Beijing 100871, China}
\author{Jakub Zakrzewski}
\email{jakub.zakrzewski@uj.edu.pl}
\affiliation{Institute of Theoretical Physics, Jagiellonian University in Krak\'ow,  \L{}ojasiewicza 11, 30-348 Krak\'ow, Poland }
\affiliation{Mark Kac Complex
Systems Research Center, Jagiellonian University in Krakow, Krak\'ow,
Poland. }

\date{\today}% It is always \today, today,

%\pacs{03.67.Lx, 42.50.Dv}% PACS, the Physics and Astronomy
                              
\begin{abstract}
We show that the presence of a harmonic trap may in itself lead to many-body localization for cold atoms confined in that trap in a quasi-one-dimensional geometry.
Specifically, the coexistence of delocalized phase in the center of the trap with localized region closer to the edges is predicted with the borderline dependent on 
the curvature of the trap. The phenomenon, similar in its origin to Stark localization, should be directly observed with cold atomic species. We discuss both the spinless and
the spinful fermions, for the latter we address Stark localization at the same time as it has not been analyzed up till now.
\end{abstract}

\maketitle

For a long time, it has been believed that many-body systems tend to thermalize as expressed by eigenstate thermalization hypothesis \cite{Deutsch91,Srednicki94}. The many-body
localization (MBL) phenomenon (for reviews see  \cite{Huse14, Nandkishore15, Alet18, Abanin19}))
is a direct counterexample -- for a sufficiently strong disorder, the system preserves the memory of its initial state. However, 
recent examples of many-body quantum so called scar states \cite{Turner2018, James2019}, Hilbert-space fragmentation \cite{Khemani20, Sala20,Gromov20,Feldmeier20}, and lack-of-thermalization in gauge theories \cite{brenes2018, Chanda20,Magnifico19} reveal  strong non-ergodic behaviors even in the absence of disorder.
 Another example considers Stark localization -- where the presence of a static electric field resulting in a tilt in the many body system may lead to localization \cite{vanNieuwenburg19, Schulz19}.

It seems that the more many-body physics is explored the less ergodic the many-body dynamics turns out to be. The present work provides another example of such a situation. 
We consider finite system sizes only.  Such systems are directly amenable to experimental studies \cite{Schreiber15,Smith16,Roushan17,Luschen17,Silevitch17,Wei18,Xu18,Guo19,Lukin19,Rispoli19}.
In this way we also stay away from a current
 vivid debate about the very existence of MBL in the thermodynamic limit \cite{Suntais19,Abanin19z,Sierant20,Panda19,Sierant20poly}.
We consider one-dimensional (1D) chains with chemical potentials {quadratically} dependent on position. Such a situation is quite common in quasi-1D situations realized in optical lattices
\cite{Fallani07,Zakrzewski09}, where a tight confinement in directions perpendicular to a chosen one is due to illumination by strong laser beams with  gaussian transverse profiles. Those profiles may be well approximated as a harmonic trap along the considered direction \cite{Fallani07}. Thus we shall consider models with  Hamiltonians being
\begin{equation} 
H=H_0+H_{trap} = H_0 + \frac{A}{2}\sum_{l=-L/2}^{L/2}l^2,
\label{genham}  
\end{equation}
where $A$ is the curvature of the harmonic trap and $l$ is the site index (we assume unit spacing between sites of the chain). $H_0$ is the model Hamiltonian considered, which may represent the Heisenberg chain (equivalent to interacting spinless fermions), bosons represented by Bose-Hubbard model, or spinful fermions with Hubbard Hamiltonian. In contrast to the study of \cite{Schulz19}, where small quadratic potential has been considered on top of {the dominant} uniform linear potential, we consider the effect of harmonic trap alone, {which as we shall show acts differently}  in different parts of the system. {Let us mention that such a model, for sufficiently big curvatures, may lead to a local quadruple conservation
\cite{Khemani20} -- thus it belongs to {a class of} fracton systems where generically slow subdiffusive approach to thermalization is expected \cite{Gromov20,Feldmeier20}. For completeness, we mention that very slow dynamics was predicted also for harmonic trap quenches for noninteracting case \cite{Schulz16}. The results presented here are limited to moderate time scales where we observe no traces of the very slow thermalizing dynamics expected in the harmonic trap \cite{Gromov20}. }

 \begin{figure}
 \includegraphics[width=\linewidth]{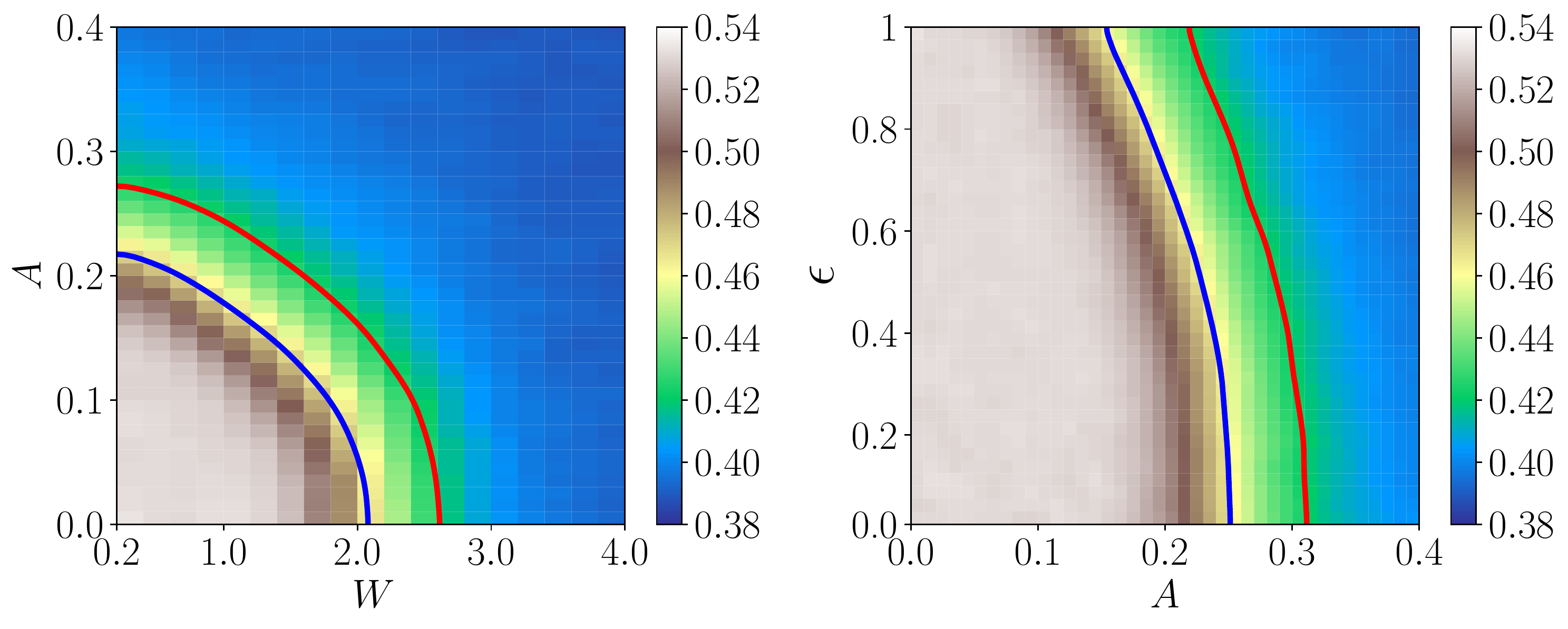}
  \caption{ {Mean gap ratio $\average r$ for the Heisenberg chain of $L=16$ sites. 
  Left: $\average r$ as a function of the disorder amplitude and curvature $A$.   In the absence of external curvature $A=0$
  the crossover to localized regime occurs for  $W\approx 2.3$ for such a small system. Above $A=0.3$, system seems, {on the basis of mean gap ratio value,} localized for arbitrary $W$. Right:  $\average r$ as a function of scaled energy $\epsilon$ and curvature $A$ at disorder strength $W=0.5$. 
{The rescaled energy is defined as $\epsilon=(E-E_{\min})/(E_{\max}-E_{\min})$, where $E_{\min}$ ($E_{\max}$)
is the energy of the ground (highest excited) state and $E$ denotes the energy of the corresponding eigenstate.}   
The blue and red lines are contours for $\average r \approx 0.46$ and $0.42$ respectively and serve as the guide to the eyes for distinguishing {apparently} localized and ergodic phases.
    \label{rbarheis} \vspace{-0.25cm}
 }}
\end{figure}

\noindent {\it Heisenberg spins or spinless fermions.--}
As the simplest possible model, first we consider  the Heisenberg chain, where $H_0$ becomes
\begin{equation}
 H_0= J\sum_{l=-L/2}^{L/2-1} \ \vec{S}_l \cdot \vec{S}_{l+1} + \sum_{l=-L/2}^{L/2}  h_l S^z_l,
 \label{eq: XXZ}
\end{equation}  
 with $ \vec{S}_l $'s being  spin-1/2 operators and $h_l$ is a diagonal disorder (a magnetic field along $z$-axis) drawn from random uniform distribution in $[-W,W]$ interval. We set $J=1$ to be the unit of energy.
 The harmonic trapping potential in this case is given by $H_{trap} = \frac{A}{2} \sum_{l=-L/2}^{L/2}  l^2 S^z_l$.
The Hamiltonian \eqref{eq: XXZ} is a paradigmatic model for MBL studies \cite{Luitz15,Sierant19b} - it maps to an interacting chain of spinless fermions via Jordan-Wigner transformation. A typical random matrix theory (RMT) based measure is the gap ratio defined as a minimum of the ratio of consecutive 
 level spacings, $r_n= \min \{ \frac{s_{n+1}}{s_n}, \frac{s_n}{s_{n+1}} \}$ with $s_n=E_{n+1}-E_n$ and {$E_n$ being the energy eigenvalue}. The mean gap ratio is $\average{r}\approx 0.53$ for delocalized system, well described by Gaussian orthogonal ensemble (GOE), while  $\average{r}\approx 0.38$
 for Poisson spectra characteristic for integrable, localized cases \cite{Oganesyan07}. 
  We find that for a sufficiently large curvature $A$ the  {mean gap ratio takes the latter value} regardless
 of the disorder amplitude (see Fig.~\ref{rbarheis} for $L=16$). The figure resembles that observed for Stark localization \cite{vanNieuwenburg19}.

\begin{figure}
  \includegraphics[width=\linewidth]{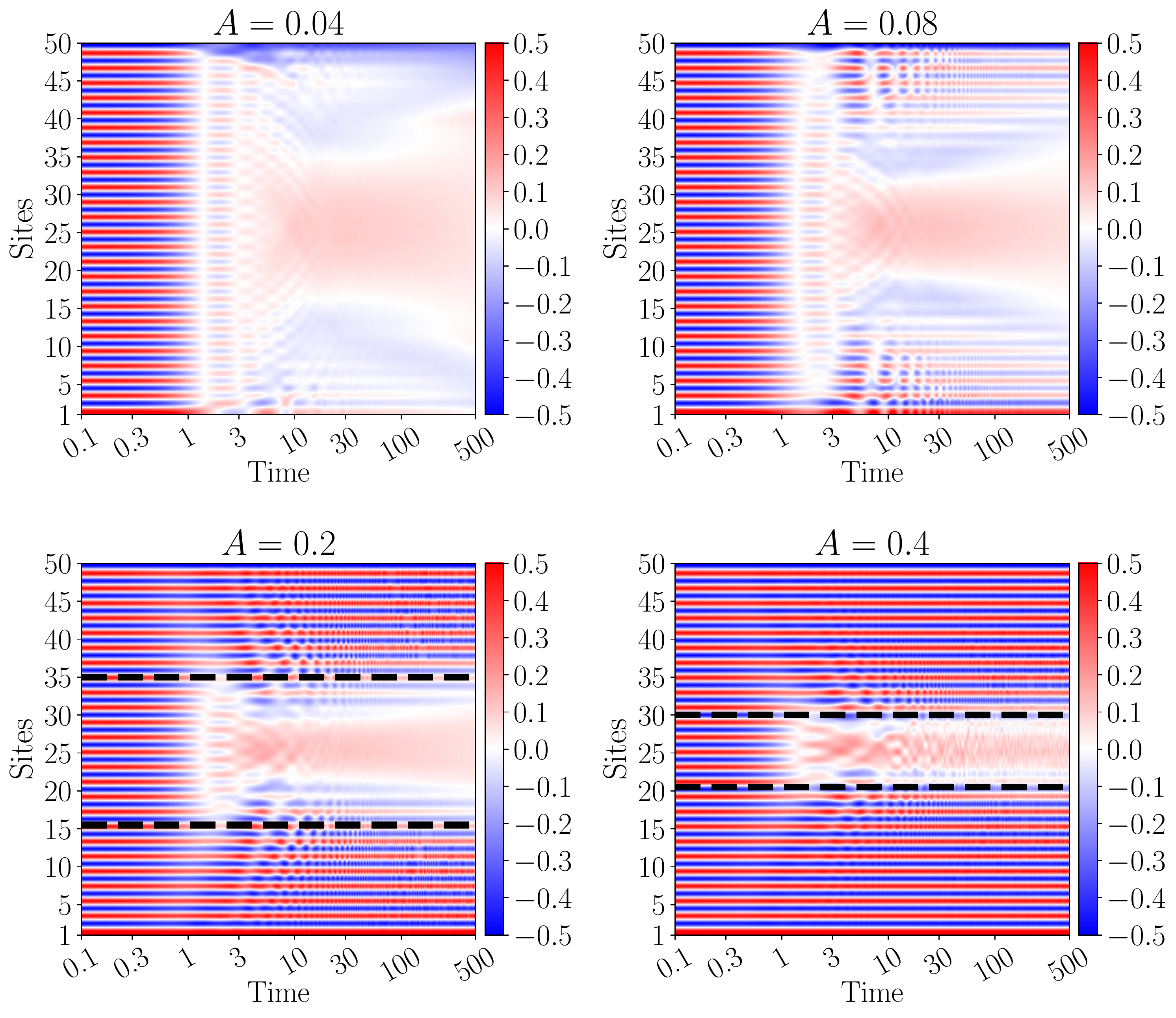}
   \caption{ Site-observed spin dynamics, as measured by the local expectation values $\braket{S^z_l}$ for $L=50$ Heisenberg chain with no disorder. For small $A$ the system ``thermalizes'' and initial spin-wave configuration is destroyed by interactions. For larger curvature, one clearly observes the coexistence of delocalized (in the center) and 
  localized regions (at the edges). The black dashed lines give the border of localization as given by Stark localization prediction with $F_c \approx 2$ \cite{vanNieuwenburg19}.
    \label{largheis} 
 }
\end{figure}

To get insights into the physics observed, let us consider the time dynamics. We prepare the chain in the separable state with every second spin being up and down respectively as $\ket{\uparrow,\downarrow,\uparrow,.., \downarrow}$ and observe whether this spin-wave arrangement is preserved in time evolution. For small disorder and small $A$ the system thermalizes (upper row in Fig.~\ref{largheis}). For larger $A$ different picture emerges -- while at the center of the chain delocalization still occurs, at a sufficient distance from it we observe preservation of the initial spin texture - i.e. localization.

\begin{figure}
  \includegraphics[width=\linewidth]{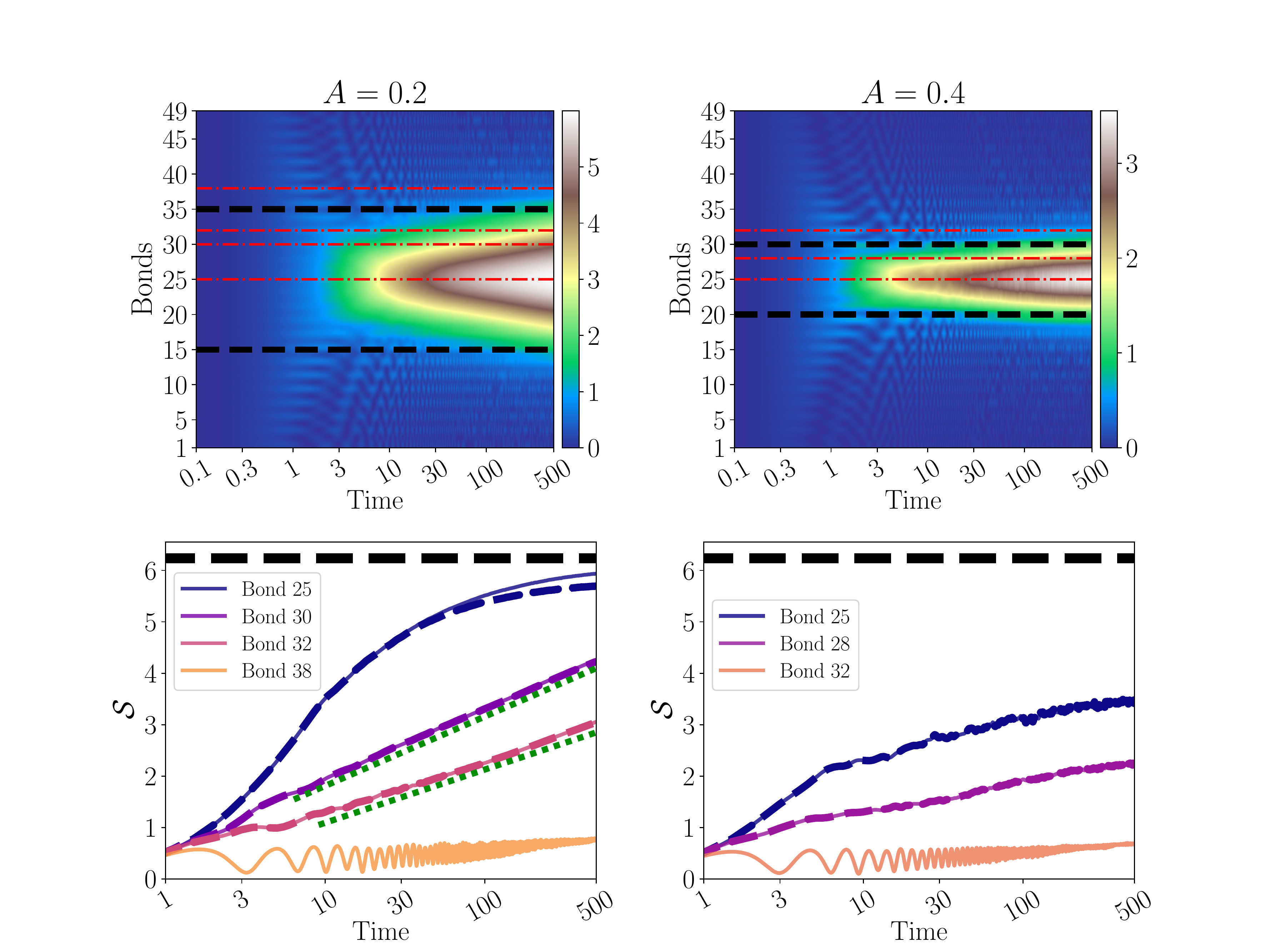}
   \caption{Time dynamics of the entanglement entropy for $A=0.2$ (left column) and $A=0.4$ (right column) for the Heisenberg chain with the initial spin-wave state and without any disorder. 
Top row: Spatial profiles of time evolved entropy measured across every bond.
The entanglement grows rapidly in the central region remaining relatively low at the edges behind borders (depicted by black dashed lines) given by the Stark localization criterion.
Bottom row: The growth of entanglement entropy with time measured across selected bonds (marked by red dash-dot lines in the top row). 
Solid lines show the results for MPS bond dimension $\chi_{max} = 512$, while thicker dashed lines are for $\chi_{max} = 384$.
The black dashed lines show the maximum allowed value $\ln 512$ of entanglement entropy by the MPS ansatz with $\chi_{max} = 512$ and
the green dotted lines in the bottom left figure are straight lines in the logarithmic scale and are there as the guide to the eyes.
    \label{entheis}
 }
\end{figure}

One can easily, {\it a posteriori}  explain this phenomenon. For a given distance $l_0$ from the center the local static field can be expressed as 
$F = \frac{\partial}{\partial l_0} \left(\frac{A}{2} l_0^2\right) = l_0 A$.
If this local field exceeds the border of Stark localization \cite{Schulz19, vanNieuwenburg19} -- the part of the system localizes, while the region close to the center remains extended. 
{Therefore, unlike the usual Stark localization, one can always find localized regions for any finite values of $A$ for large enough systems under harmonic trapping potential.}
The dashed lines in Fig.~\ref{largheis} give the Stark localization border, as predicted in \cite{vanNieuwenburg19} to be $F \approx 2$, which nicely fits numerical data.

While Fig.~\ref{largheis} clearly shows the coexistence of localized (close to edges) and delocalized (in the center of the trap) regions, this finding seems to be in contradiction with the mean gap ratio data of Fig.~\ref{rbarheis}, which indicates that  $\average r$ takes the value close to Poissonian-like for $A=0.4$. Such a value may correspond to a fully localized case, but also to a superposition of independent spectra. {Therefore, a logical consequence is that the eigenstates are either localized close to the edges or extended over the central region, 
such that
the mean gap ratio value comes as a result of the superposition of three independent spectra, only two of them being localized.}

 The simulations of time evolution are performed using time-dependent variational principle (TDVP) algorithm using matrix product states (MPS) ansatz \cite{haegeman_prl_2011, koffel_prl_2012, haegeman_prb_2016, paeckel_aop_2019} . More specifically, we use a \textit{hybrid} variation of the TDVP scheme mentioned in \cite{goto_prb_2019, paeckel_aop_2019, Chanda20, Chanda20many}, where we first use two-site version of TDVP to dynamically grow the bond dimension up to a prescribed value, say $\chi_{max}$. When the bond dimension in the MPS is saturated to  $\chi_{max}$, we shift to the one-site version to avoid any errors due to truncation in singular values that appears in two-site version \cite{paeckel_aop_2019, goto_prb_2019}. The final results are produced with $\chi_{max} = 512$, so that the maximum allowed value of the entanglement entropy at any given bond in the bulk of the system is $\ln \chi_{max} = \ln (512)$.

{Instead of spin profiles, one may look at the entanglement entropy growth in time (see Fig.~\ref{entheis}). We see that entropy grows rapidly in the central region, while remaining low in the localized parts. }
In the delocalized center the entropy grows fast, and therefore, the simulations in this region may not be accurate. However, as we move towards the boundaries of the system, the results with $\chi_{max}=512$ become `exact', even within the delocalized region. We confirm this by performing the same simulations with $\chi_{max} = 384$. The bottom row of Fig.~\ref{entheis}  shows such a comparison of results with $\chi_{max} = 512$ and $384$ for $L=50$ Heisenberg chain with $A=0.2$ and $0.4$. Here, we only compare bonds that are in the delocalized part, as we always get converged results in the localized regions even for $\chi_{max} = 384$. {The comparison of TDVP data with numerically exact results obtained using Chebyshev expansion of the time evolution operator \cite{Tal-Ezer84,Cheby91,Fehske08}  for $L=20$ is presented in \cite{suppl}.}

Quite surprisingly, the entanglement entropy can also show logarithmic growth in time, even in the delocalized regions when the effect of trapping potential becomes strong. For example, in case of $A=0.2$, entropy the central bond grows rapidly in time and approaches the maximum allowed value by the MPS ansatz ($\ln(512)$ in this case. On the other hand, the entropy in the bonds 30 and 32 shows a logarithmic growth, despite being on the delocalized side of the system. The entanglement growth in case of $A=0.4$ is greatly modified by the harmonic trap even in the central bond. This is indeed very unusual dynamics where delocalized parts behave as systems showing MBL in terms of entropy growth.
{The plausible explanation of this behavior comes from the fact that  entanglement entropies between nearby sites cannot 
differ much due to the local Hilbert space dimension being equal to 2 for spins. Thus logarithmic slow growth in time in localized region affects also sites being the close neighborhood of the border between localized and delocalized sites.}

{One important point to mention is that} to observe clear signatures of MBL for the pure linear lattice tilt, either a small disorder or a slight curvature has to be added to the potential \cite{Schulz19,vanNieuwenburg19} to avoid
thermalization due to fracton dynamics \cite{Taylor19}. For our harmonic potential the local field changes from site to site, that apparently suffices to avoid subdiffusive thermalization. Another interesting point is that the observed results do not warrant a finite-size scaling, as for a fixed value of $A$ the central region remains almost invariant and simply additional localized sites are added in the edges with increasing system size. For illustration, we put the results for $L=20$ in \cite{suppl}.

\begin{figure}
  \includegraphics[width=\linewidth]{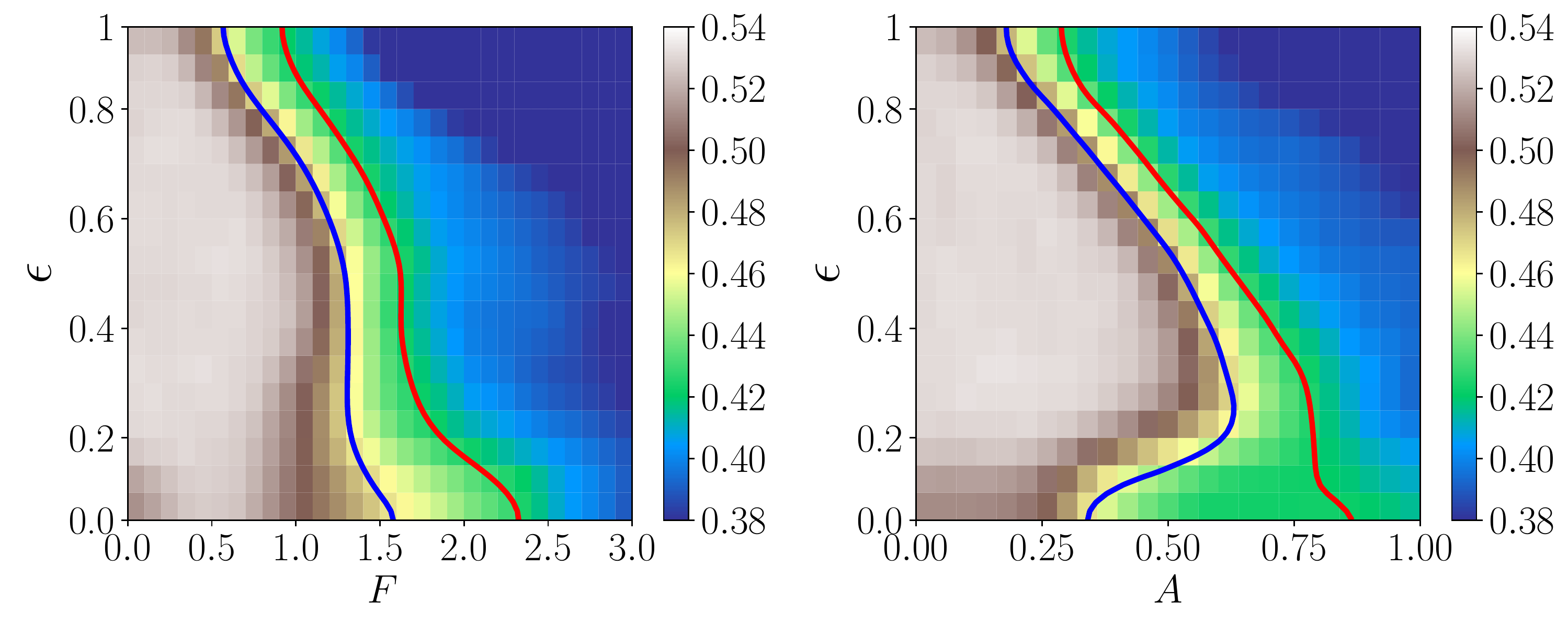}
   \caption{Left: Gap ratio statistics for a static electric field problem for spinful fermions with interaction strength $U=1$ as a function of scaled energy $\epsilon$ and static field amplitude $F$ for disorder strength $W=0.5$. The plot corresponds to 6 fermion-system in $L=12$ lattice sites
   (quarter filling) with number of up and down fermions being equal. The transition to Poisson statistics is smooth and extended over a range of static field $F$ values. Stark localization is observed for sufficiently large $F$. 
   Right: The gap ratio statistics for spinful fermions in the harmonic trap for disorder strength $W=0.5$.
Other parameters are   same  as in the left panel. The blue and red lines in both the figures are contours for $\average r \approx 0.46$ and $0.42$ respectively and serve as the guide to the eyes for distinguishing localized and ergodic phases.
    \label{stark}
 }
\end{figure}

\begin{figure}
   \includegraphics[width=\linewidth]{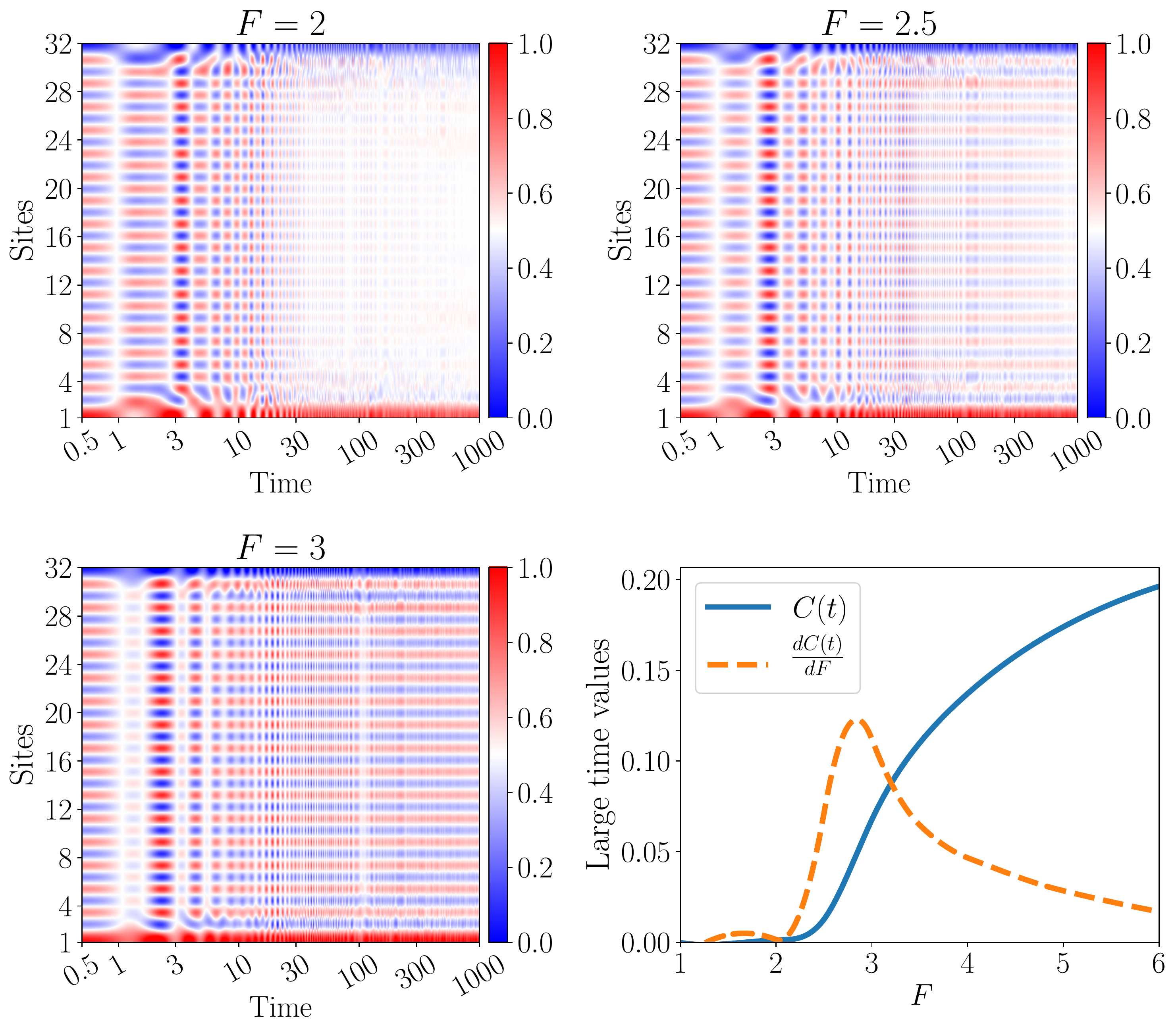}
   \caption{ Top row and bottom left panel: Time dynamics of spinful fermions for the initial staggered density-wave state without any disorder for three static field amplitude, $F$, across the crossover. The localization is complete for
 $F=3$, while for intermediate field values a partial localization is observed with some fraction of particles accumulating near the bottom of the effective potential. 
 Bottom right panel: Time evolved density correlator $C(t)$ at large times as a function of $F$. We discard 4 sites from the boundaries to minimize their effects.
 We take the inflection point of the curve at $F_c \approx 2.8$ as a critical field amplitude. All data are for $L=32$ obtained using TDVP algorithm.
  \label{starktime}
 }
\end{figure}

\noindent {\it Spinful fermions: Stark localization.--} 
Since the interactions between spinless fermions are hard to realize experimentally {in a standard cold atoms in optical lattice setting}, we consider the spinful case, represented by the Hubbard model, as in experiments \cite{Schreiber15,Luschen17}. The curvature free part $H_0$ is
\begin{equation}
    {H}_{0} = - J\sum_{l, \sigma} \left(\hat{c}^{\dagger}_{l \sigma} \hat{c}_{l+1 \sigma} 
		   + {\rm h.c.}\right) 
		   + U \sum_{l} \hat{n}_{l\uparrow} \hat{n}_{l\downarrow} 
		   + \sum_{l,\sigma} h_{l} \hat{n}_{l\sigma},
\label{model}
\end{equation}
with $l \in [-L/2, L/2]$.
As before, we set $J=1$ to be the unit of energy and consider $U=1$ throughout this communication unless otherwise stated.
Let us first consider the Stark localization problem under linear potential as it was only addressed for spinless fermions \cite{Schulz19,vanNieuwenburg19} until now. Thus, we add to the Hamiltonian a tilt term $F \sum_l l (\hat{n}_{l \uparrow} + \hat{n}_{l \downarrow})$
and analyze the gap ratio statistics.
The corresponding $\average r$ statistics is shown in Fig.~\ref{stark}(a) for $L=12$ quarter-filled chain (i.e., $N_{\uparrow} = N_{\downarrow} = L/4$). {To obtain this plot we break the SU(2) symmetry of the Hamiltonian by adding a local magnetic field to the Hamiltonian via the term $H_{break}=B(n_{L/2\uparrow}-n_{L/2\downarrow})$ with $B=0.5$ following the prescription and discussion in \cite{Mondaini15}.} We observe that, in comparison to spinless fermions \cite{vanNieuwenburg19}, the crossover seems quite broad, possibly due to the small system size taken. 
To get more precise critical value of $F$, we consider time evolution of staggered density-wave state $\ket{\uparrow, 0, \downarrow, 0, \uparrow, 0, \downarrow,...}$ for larger system-sizes and measure the density correlation
$C(t) = D \sum_l \left(\bar{n}_l(t) - \rho\right)\left(\bar{n}_l(0) - \rho\right),$
where $\bar{n}_l(t) = \braket{\hat{n}_{l \uparrow} + \hat{n}_{l \downarrow}}$, $\rho$ is the average number of particles per site, and the constant $D$ is chosen so that  $C(0) = 1$. The illustration of such time dynamics for $L=32$ sites system is shown in Fig. \ref{starktime}.
We observe that for the high field value, e.g., $F=3$, the localization is almost complete (bottom left panel). On the other hand, at lower fields, we observe a partial localization (as revealed also by standard local densities $\bar{n}_l(t)$), see top row of Fig. \ref{starktime}). The bottom right panel of Fig. \ref{starktime}  shows large time values of the density correlator $C(t)$ and its derivative $\frac{d C(t)}{d t}$. We approximate critical $F_c$ from the inflection point of $C(t)$, as obtained from the maximal value of its derivative,  to be $F_c \approx 2.8$ for disorderless scenario.

\begin{figure}
  \includegraphics[width=\linewidth]{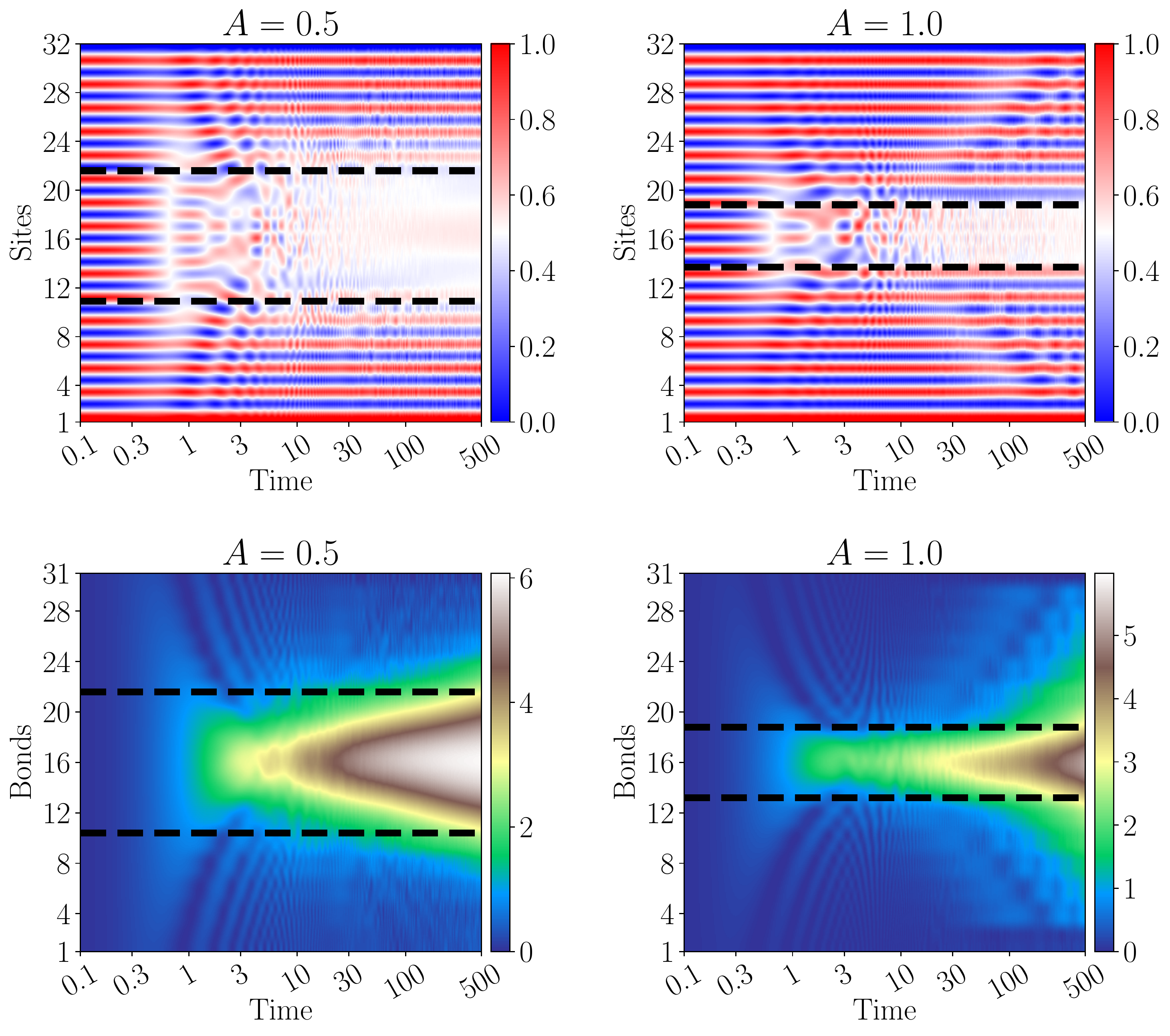}
   \caption{ Top row: Time-evolved density profile of spinful fermions for the initial staggered density-wave state under harmonic potential with no disorder.
   Bottom row: Time dynamics of entanglement entropy measured across every bond for the same systems as in the top row.
All the figures are for quarter-filled $L=32$ chain. Black dashed lines gives the physical border of localization as predicted by the Stark localization with $F_c \approx 2.8$.
    \label{l32dens}
 }
\end{figure}

\noindent {\it Spinful fermions: Localization under harmonic trap.--}
Having established the estimate for the critical field amplitude corresponding to the crossover to localized phase, we may turn again to the harmonic confinement case. Thus we again consider \eqref{genham} now for spinful fermions \eqref{model}, with $H_{trap} = \frac{A}{2}\sum_{l=-L/2}^{L/2} l^2 (\hat{n}_{l \uparrow} + \hat{n}_{l \downarrow})$. 
The right panel of Fig.~\ref{stark} shows level spacing statistics for quarter-filled $L=12$ chain in such case and 
Fig.~\ref{l32dens}  depicts the time dynamics of the density profile 
for different curvatures of the harmonic potential with the staggered density-wave state being the initial one. We observe that, as in the spinless fermions case, while in the center of the trap apparent fast ``thermalization'' occurs, closer to the edges the effective electric field coming from the curvature of the trap leads to localization. The dashed lines give the estimate of the threshold assuming $F=l_0 A$ condition with $F_c = 2.8$ from the previous analysis. 

We may also analyze the time evolved entanglement entropy in different regions. 
While in the center of the trap the entropy grows linearly with time and soon saturates due to insufficient bond dimension of the MPS ansatz rendering the results in this region not accurate, the entropy beyond the localization boundary given by $l_0 > F/A$ (and symmetrically for negative $l_0$) seems to grow logarithmically providing a further evidence for many-body localization in the outer regions.

\begin{figure}
  \includegraphics[width=\linewidth]{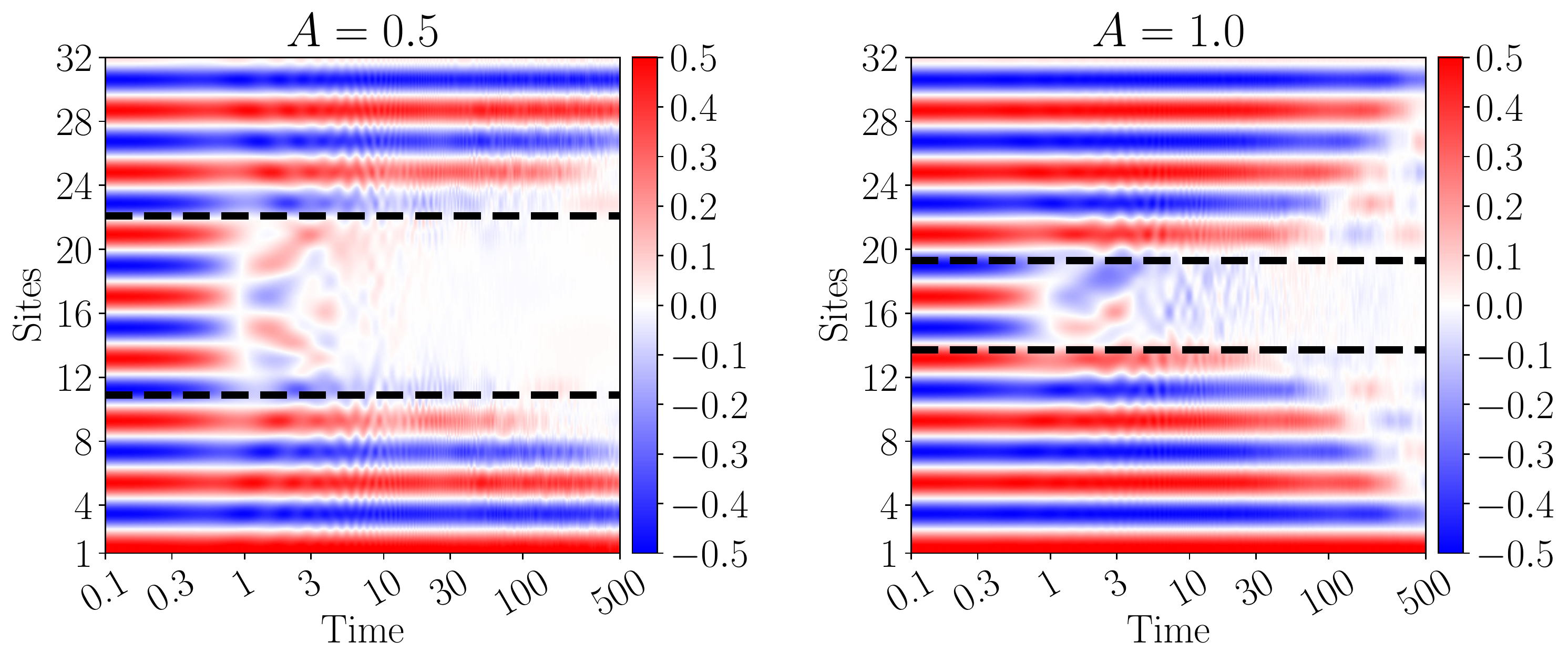}
   \caption{Time-evolved profile of local spin operator $S^z_l = (\hat{n}_{l \uparrow} - \hat{n}_{l \downarrow})/2$
   of spinful fermions for the initial staggered density-wave state under harmonic potential with no disorder.
Other specifications are same as in Fig~\ref{l32dens}.
 }
    \label{l32spins}
\end{figure}

The corresponding spin dynamics is also interesting. As in the standard MBL case, we observe a subdiffusive decay of initial spin configuration for the initial staggered density-wave state, characteristic
of the remaining SU(2) symmetry of the problem \cite{Prelovsek16,Sroda19, Zakrzewski18}. Visualization of this effect {can be seen from the profile of spin degrees of freedom in Fig.~\ref{l32spins}, where a slow spreading of the delocalized region in the spin sector at later times can be observed.}

\noindent {\it Conclusions.--} We have shown that many-body localization behavior can be observed in the presence of the harmonic trap and in the absence of the disorder. The
effect is due to a local static field that induces, for sufficient curvatures, Stark localization as recently shown for spinless fermions \cite{Schulz19,vanNieuwenburg19,Taylor19} and announced in the spinful case \cite{bloch20}.
Since the effect has a lower bound on the curvature, the central region of the trap remains delocalized. Thus a harmonic trap makes a possible realization of a very interesting situation -- coexistence of delocalized and MBL phases in a single system. %Let us note that these regions do not suffer from avalanche effects \cite{Thiery18} at least on the experimental time scales considered. 
{Let us stress that on the experimentally relevant time-scales considered by us, we do not observe any traces of the slow subdiffusive thermalization predicted due to fracton hydrodynamics \cite{Gromov20}.} Finally, let us also mention that the harmonic trap may play some role in the experiments on MBL {performed in optical lattices} (e.g. \cite{Schreiber15,Luschen17}) as the residual trap due to Gaussian beam profiles is most probably present there. While in these experiments disorder induced effects play a dominant role the residual harmonic-like trap may affect the details of the time dynamics for
large systems. This aspect is a subject of a current research. {Finally, we note that the effect is not limited to harmonic trap but can be generalized to arbitrary potentials with  non-vanishing first order derivatives.}

 \begin{acknowledgments} 
\noindent {\it Acknowledgments.--} 
The numerical computations have been possible thanks to  High-Performance Computing Platform of Peking University as well as PL-Grid Infrastructure.
The TDVP simulations have been performed using ITensor library (\url{https://itensor.org}).
This research has been supported by 
 National Science Centre (Poland) under projects  2017/25/Z/ST2/03029 (T.C.) and  2016/21/B/ST2/01086 (J.Z.)
 \end{acknowledgments}

\bibliographystyle{apsrev4-1}
\bibliography{ref_2006.bib} 
\end{document}

\title{Supplementary material for \\
``Coexistence of localized and extended phases:\\
Many-body localization in a harmonic trap''}
% Force line breaks with \\

\author{Titas Chanda}
\email{titas.chanda@uj.edu.pl}
\affiliation{Institute of Theoretical Physics, Jagiellonian University in Krak\'ow,  \L{}ojasiewicza 11, 30-348 Krak\'ow, Poland }

\author{Ruixiao Yao}
\affiliation{School of Physics, Peking University, Beijing 100871, China}
\author{Jakub Zakrzewski}
\email{jakub.zakrzewski@uj.edu.pl}
\affiliation{Institute of Theoretical Physics, Jagiellonian University in Krak\'ow,  \L{}ojasiewicza 11, 30-348 Krak\'ow, Poland }
\affiliation{Mark Kac Complex
Systems Research Center, Jagiellonian University in Krakow, Krak\'ow,
Poland. }

\date{\today}% It is always \today, today,

%\pacs{03.67.Lx, 42.50.Dv}% PACS, the Physics and Astronomy
                              
\begin{abstract}
We provide  additional evidences for the phenomena reported in the main text.
\end{abstract}

\maketitle

\begin{figure}[b]
 \includegraphics[width=\linewidth]{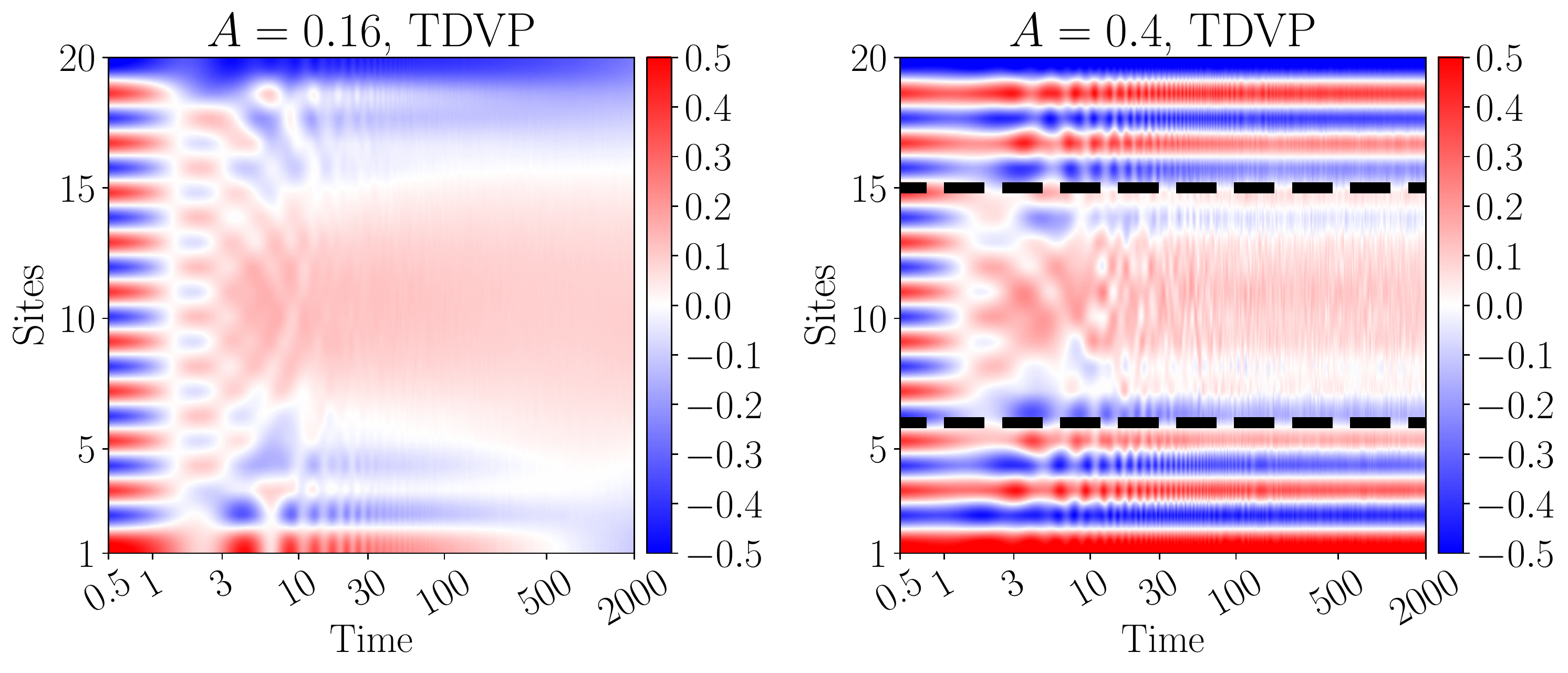} \\
  \includegraphics[width=\linewidth]{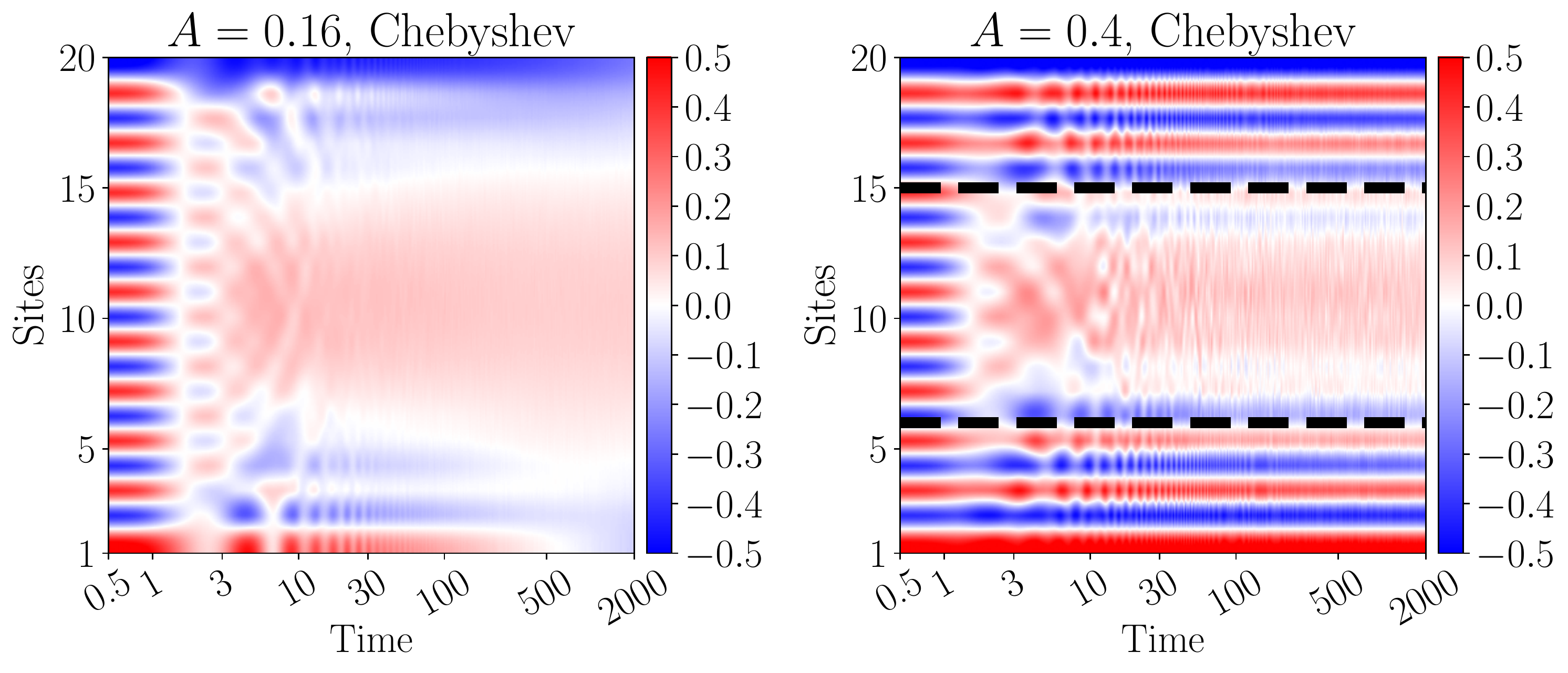}
 \caption{ Site-observed spin dynamics for small for $L=20$ Heisenberg chain without any disorder for the initial spin-wave state $\ket{\uparrow, \downarrow, \uparrow, \downarrow, ...}$.  We observe similar behavior as reported in the main text for $L=50$.
 The dashed line indicate location of the local critical electric field, $F_c = l_c A$ with $l_c$ being the site number corresponding to the onset of Stark localization.
 Top row: results from TDVP simulation. Bottom row: exact results using Chebyshev expansion method.
   \label{L20heis} 
 }
\end{figure}

\begin{figure}
 \includegraphics[width=\linewidth]{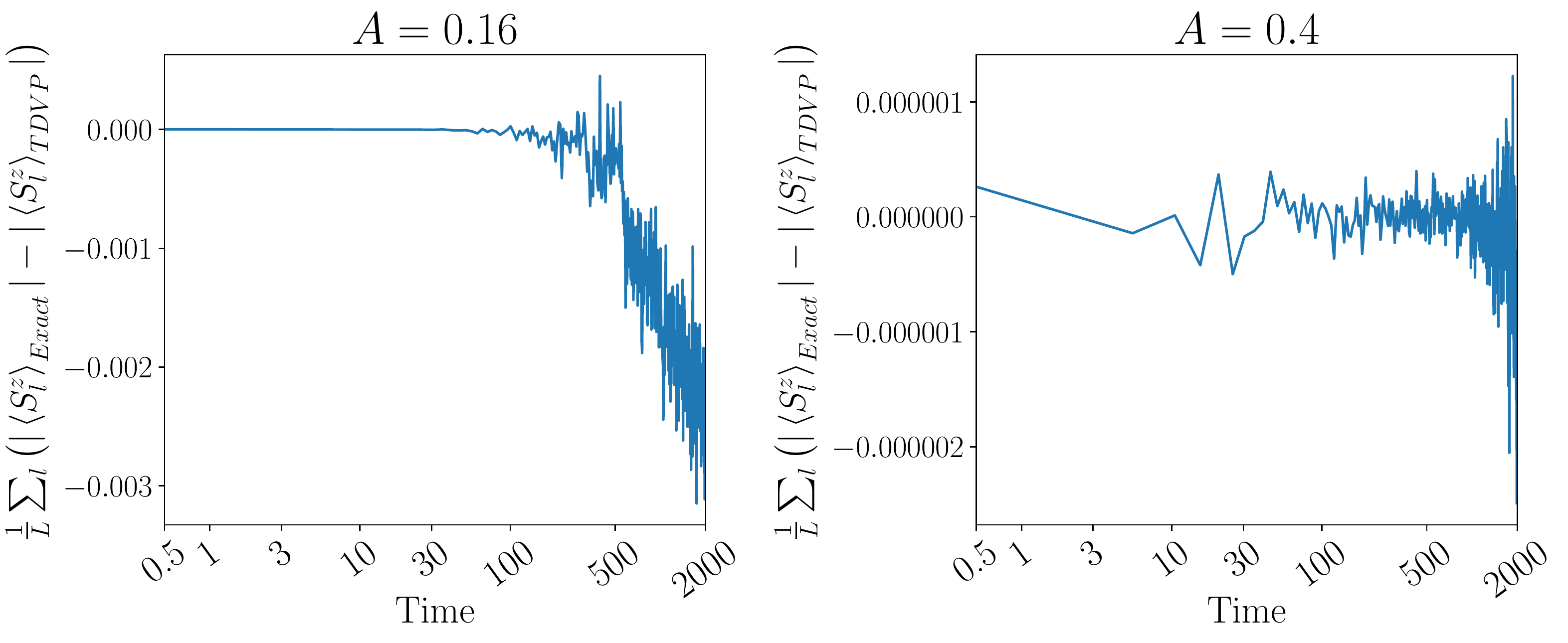} 
 \caption{ Site-averaged errors of magnetizations in TDVP data for $L=20$. Clearly, `hybrid' TDVP method used in this study can provide very accurate results even at longer times, especially for larger values of $A$.
  \label{L20heis_errr} 
 }
\end{figure}

\subsection{Heisenberg spins or spinless fermions}
While in the main text we have shown the localization in a harmonic trap for $L=50$, it is fruitful to consider smaller system sizes. Fig.~\ref{L20heis} shows the effect of the harmonic trap for two different curvatures $A$. For too small curvature, no localization effect is observed. With increasing $A$ the system starts to localized at the edges, when the local electric field $F=l A$ (where $l \in [-L/2,L/2]$ is the site number) exceeds the Stark localization value $F_c \approx 2$  \cite{vanNieuwenburg19}. {In Fig. \ref{L20heis}, we also compare the spin profiles coming from TDVP simulations with the exact data from Chebyshev expansion of the time evolution operator \cite{Tal-Ezer84,Cheby91,Fehske08}. Clearly, the TDVP simulations can faithfully capture the time dynamics in these systems, which can be further verified in Fig. \ref{L20heis_errr}, where we plot site-averaged errors in magnetizations that exists in TDVP simulations.}

\begin{figure}
  \includegraphics[width=\linewidth]{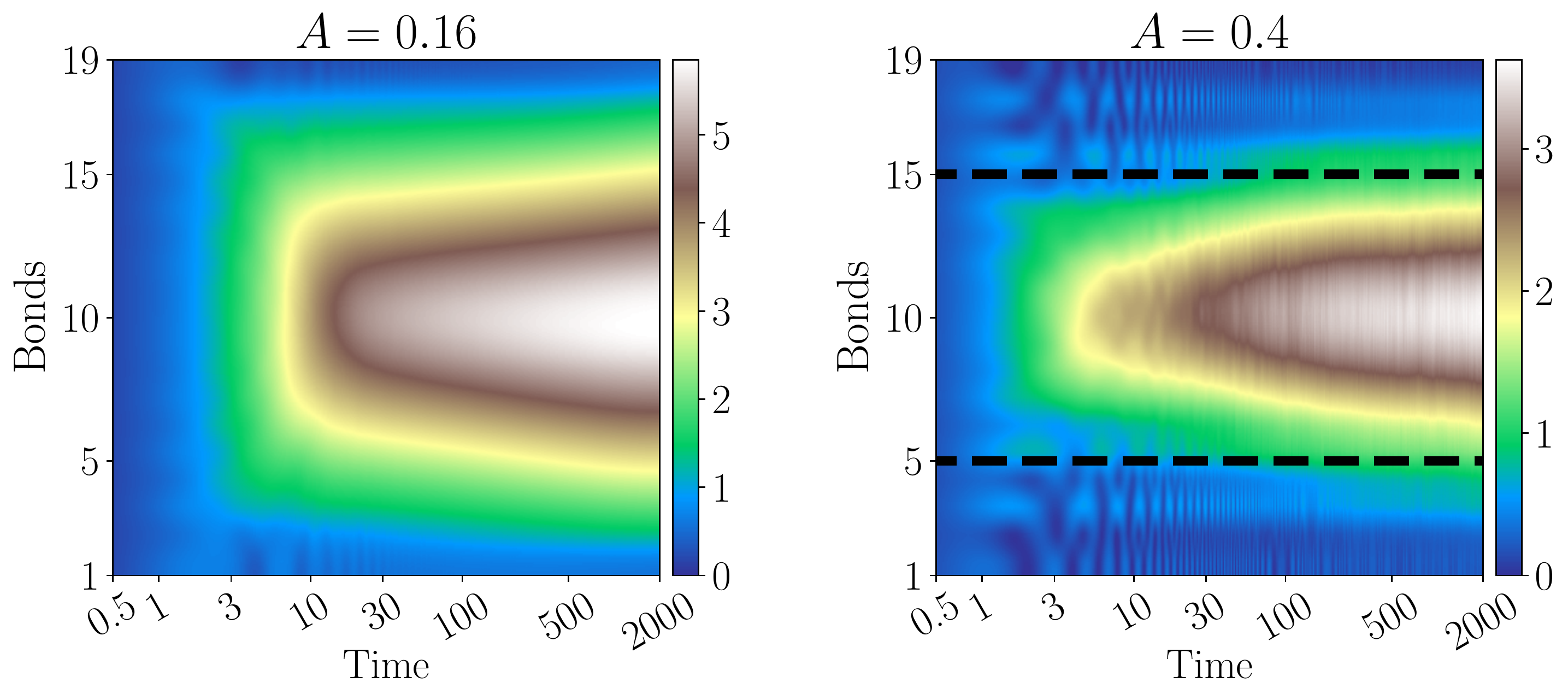}
   \caption{Time dynamics of the entanglement entropy for $L=20$ chain measured across every bond. Entropy grows rapidly in the central delocalized region, remaining relatively low at the edges behind borders given by Stark localization criterion.
    \label{entL20heis}
 }
\end{figure}

In Fig.~\ref{entL20heis}, we show the time-evolved pattern of entanglement entropy for the same. The entropy grows rapidly in the central region which is apparently delocalized, and grows slowly in the localized sites close to the edges of the system. For smaller values of $A$ (e.g., 0.16), the entropy follows `volume-law' growth with the size of the bipartition in the time-evolved asymptotic states, while for larger values of $A$ it is concentrated only in the central region.

\begin{figure}
   \includegraphics[width=.55\linewidth]{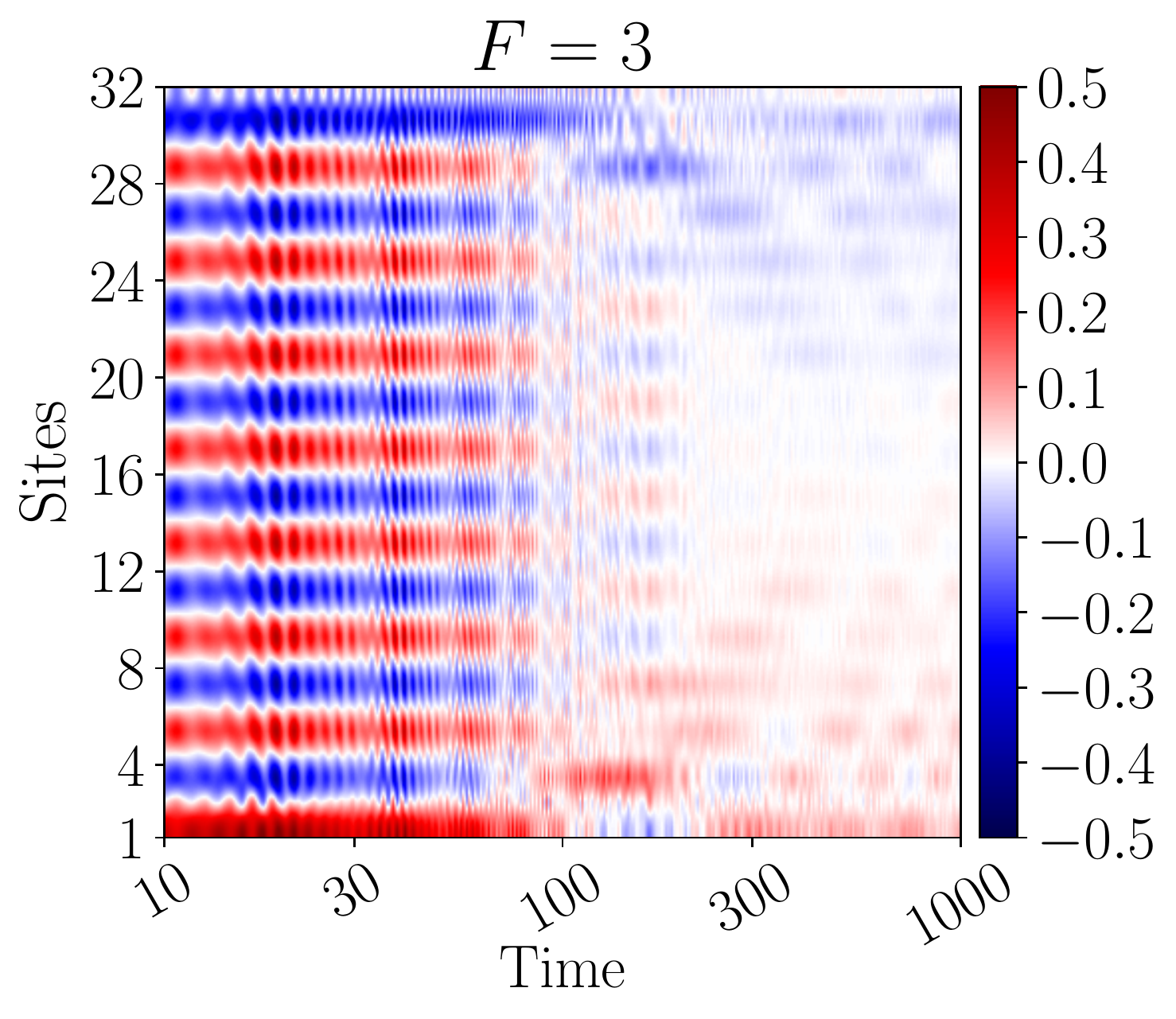}
   \caption{Time evolved profile of local spins for spinful fermions 
   without any disorder for static field amplitude $F=3$ starting from the initial staggered density-wave state.}
   \label{L32spin_ramp}
\end{figure}

\subsection{Spinful fermions} 
In the main text, we have considered Stark localization in case of spinful fermions, as described by the Hubbard Hamiltonian in presence of the tilt term $F \sum_l l (\hat{n}_{l \uparrow} + \hat{n}_{l \downarrow})$. We have analyzed the system under quarter filling with equal number of spin up and down particles, and only examined the charge (or density) sector during the time evolution via local densities $\bar{n}_l(t)$ and the density correlator $C(t)$. We now 
also consider the spin sector via the local spin operator, $S^z_l = (\hat{n}_{l \uparrow} - \hat{n}_{l \downarrow})/2$, of the same system starting from the same staggered density-wave state $\ket{\uparrow, 0, \downarrow, 0, \uparrow, 0, \downarrow, ...}$. Fig. \ref{L32spin_ramp} shows such time-evolved profile of the spin sector for quarter-filled $L=32$ chain at $F=3$ where the localization of the charge sector has been shown to be completed. However, we can easily appreciate the subdiffusion of spin sector (c.f. \cite{Prelovsek16,Sroda19, Zakrzewski18}) where the memory of the initial spin configuration starts to wear off after $t \approx 200$.

\begin{figure}
   \includegraphics[width=\linewidth]{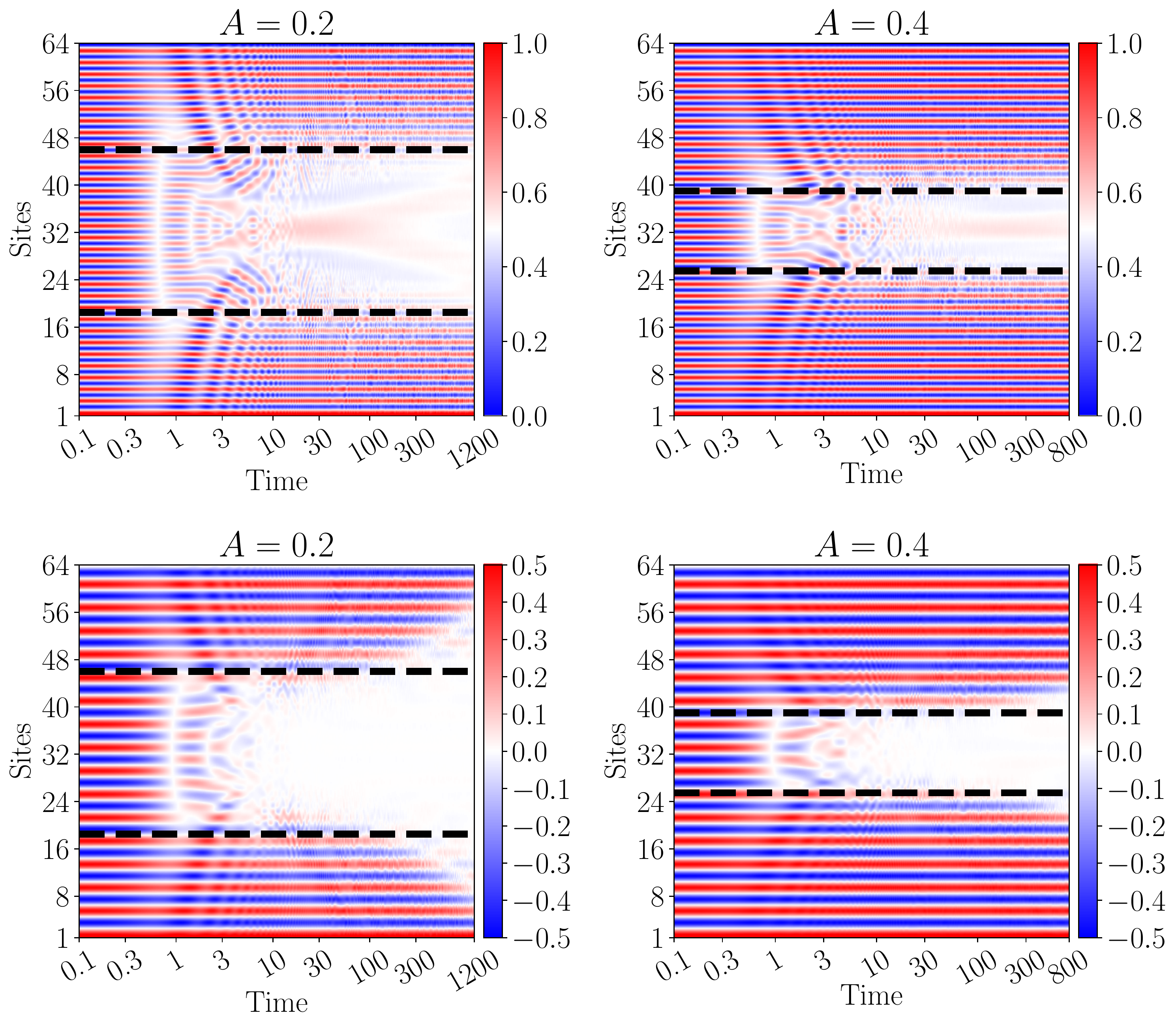}
   \caption{Time evolution of spinful fermions without any disorder under the harmonic trap for $L=64$ for two different values of curvature $A$ starting with the initial staggered density-wave state. We plot both the local densities (top row) as well as the local spins (bottom row) for same values of $A$. The black dashed lines are the predicted border between localized and delocalized regions as estimated from $F_c \approx 2.8$.
   \label{L64DensSpin}
 }
\end{figure}

Let us now move to  spinful fermions in the harmonic trap, $H_{trap} = \frac{A}{2}\sum_{l=-L/2}^{L/2} l^2 (\hat{n}_{l \uparrow} + \hat{n}_{l \downarrow})$. We immediately go to larger system sizes and present data for $L=64$. The supplement allows to present them in a full glory.
 Fig.~\ref{L64DensSpin} shows the both the 
density profile (top row) as well as the spin profile (bottom row)
for two different values of curvature $A$ with the initial state being $\ket{\uparrow, 0, \downarrow, 0, \uparrow, 0, \downarrow, ...}$. As in the case of $L=32$ reported in the main text, the predicted borders between the localized and the delocalized regions obtained from the critical field strength $F_c \approx 2.8$
match very well with the density plots for $L=64$ too. On the other hand, a slight spreading of the delocalized region in the spin sector at later times is an indicator of the subdiffusive spin dynamics \cite{Prelovsek16,Sroda19, Zakrzewski18}. Over all the qualitative features of MBL in a harmonic trap are very similar to that in Stark localization as well as in the standard disorder induced MBL.

\bibliographystyle{apsrev4-1}
%\bibliography{ref_20} 
\bibliography{ref_2006.bib}